\documentstyle[12pt]{article}

\newcommand \ga{\raisebox{-.5ex}{$\stackrel{\sim}{>}$}}
\newcommand \ea{\raisebox{-.5ex}{$\stackrel{\sim}{=}$}}

\begin{document}

%
%

\title
{Spin Gaps and Spin Dynamics
in ${La_{2-x} Sr_x CuO_4}$ and ${YBa_2 Cu_3 O_{7-\delta}}$
}

\author{
A. J. Millis\\
AT\&T Bell Laboratories\\
600 Mountain Avenue\\
Murray Hill, NJ 07974\\
and\\
H. Monien\\
Institute for Theoretical Physics\\
University of California\\
Santa Barbara, CA 93103\\
}

\maketitle

\date{\today}

\vspace{0.2cm}
\maketitle
\begin{center}
{\bf\large Abstract:}
\end{center}
{
A magnetic susceptibility which decreases
with decreasing temperature is observed in all $CuO_2$
based superconductors with less than optimal doping.
We propose that in $La_{2-x} Sr_x CuO_4$
this is due to antiferromagnetic ordering
which is prevented by the low spatial dimensionality
while in $YBa_2 Cu_3 O_{6.6}$ it
is due to the interplay between
antiferromagnetic fluctuations
within a plane and singlet pairing of electrons
between nearest neighbor planes.\\
}
Pacs: 74.65+n,76.60-k,75.30.Kz,75.10-b
\newpage
The low frequency spin dynamics of the $CuO_2$-based
superconductors are anomalous \cite{review}.
The anomaly which has received the most attention is the different temperature
dependences of the copper (Cu) and oxygen (O)
nuclear magnetic relaxation rates
$T_1^{-1}$.
The difference apparently occurs in all $CuO_2$
superconductors and has been successfully interpreted \cite{MMP}
in terms of
antiferromagnetic spin fluctuations
which increase in strength as the temperature
is lowered, and which couple much more strongly to the
Cu than the O nuclear moments.  It is still controversial
\cite{review,Rossat-Mignod,Tranquada,imai} whether
the increase in strength
is associated with a temperature dependent correlation length as
expected near an antiferromagnetic ordering transition.
A less well understood
anomaly occurs in ``underdoped'' $CuO_2$ materials
(i.e. those with less than the optimum number
of carriers) and involves a
decrease in the spin susceptibility $\chi_s$ and some relaxation rates $1/T_1T$
with decreasing temperature, as expected in systems with a gap to low-lying
spin excitations.
Several authors have proposed
that this spin gap behavior is evidence for the
existence of exotic quantum liquid phases in underdoped high
$T_c$ materials \cite{spingap}.
Most of the theoretical work on exotic quantum phases
has focussed on a possible doping induced spin-gap state
of a single $CuO_2$ plane.
In this paper we argue that the data suggest
a different physics: in $La_{2-x} Sr_x CuO_4$ at $x<0.15$
an antiferromagnetic
instability which
is prevented from developing
into true long range order by the low dimensionality
nevertheless leads to formation of
a gap in the longitudinal
spin fluctuation spectrum which leads to a decrease
(but not to zero) of $\chi_s$.
However, in $YBa_2 Cu_3 O_{6.6}$ the physics is driven by
an interplay between  tendencies towards
spin singlet pairing of electrons in adjacent $CuO_2$
planes and antiferromagnetism in
a given $CuO_2$ plane.

The ideas we present here are related to previous
work on the high $T_c$ problem.
D.\ C.\ Johnston argued early on that a Heisenberg
model with a J decreasing with increasing x explained the
$La_{2-x} Sr_x CuO_4$ data \cite{Johnston}.
Magnetic correlations are also important in the  spin bag picture of
Schrieffer, Wen and Zhang \cite{SWZ}.
Barzykin and Gorkov
\cite{Gorkov} have discussed theoretically the possiblity
of strong short-ranged magnetic correlations in $La_{2-x}Sr_xCuO_4$
and mentioned the connection of this with a $\chi _s(T)$.
Our analysis of  $YBa_2 Cu_3 O_{6.6}$
is related to previous work of Altshuler and Ioffe as discussed below
\cite{lev}.

We now turn to the data.
$\chi_s$ may be determined from the measured
bulk susceptibility
$\chi_b ~=~\chi_s~+~\chi_c ~+~ \chi_{vv}~+~\chi_{dia}$
if the core $(\chi_c)$, van\ Vleck ($\chi_{vv}$) and Landau
diamagnetic ($\chi_{dia}$) contributions are known. In $YBa_2Cu_3O_7$,
$\chi_{dia}$
has been found to be less than 1/6 of $\chi_s$ \cite{diamag}. We neglect
$\chi_{dia}$ in our
analysis. $\chi_s$ for the $YBa_2 Cu_3 O_{7- \delta}$
materials is known \cite{review,YKs}. For
$La_{2-x} Sr_x CuO_4$, however, there
are conflicting assertions in the literature
\cite{batlogg,reven,kitks,Monthoux}.
We  estimate that in $La_2CuO_4$
$( \chi_c~+~\chi_{vv} ) ~\ea~1.0 states/(eV-Cu)$ for fields parallel to the
$CuO_2$ plane and
$-1.3 states/(eV-Cu)$ for fields perpendicular to it by comparing
the measured \cite{batlogg} susceptibility (at temperatures of
order 600-800K, well above the 3d Neel ordering temperature) to the known
\cite{heis}
values for the 2d S=1/2 quantum Heisenberg model \cite{qheis}. The example of
the $YBa_2 Cu_3 O_{7- \delta}$
series \cite{review,diamag} and the near equality \cite{batlogg} of
the difference between the ab-plane and c-axis susceptiblities
of  $La_2CuO_4$ and $La_{1.925} Sr_{0.075}CuO_4$
shows that it is reasonable to assume that
$\chi_{vv}$ and $\chi_c$
do not change much
with doping; we have used this assumption to extract $\chi_s$ from $\chi_b$.

Fig.\ 1 shows $\chi_s(T)$ for some high
$T_c$ materials, and for the S=1/2 2d
 Heisenberg model with $J~=~0.13$\ eV.   Fig.\ 2 shows the copper nuclear
relaxation
rate ($1/T_1 T$) data \cite{LaSrT1,YBaT1}    It is clear that $YBa_2 Cu_3
O_{6.6}$
is different from the other materials.
Its spin susceptibility is smaller in magnitude and
larger in relative temperature dependence than all of the
others  and unlike the other materials extrapolates
to zero at
$T~=~0$.
The relaxation rates in
$La_{2-x} Sr_x CuO_4$ are
larger and more strongly temperature and doping dependent
 than in $YBa_2 Cu_3 O_{7- \delta}$.  Further, in $La_{2-x} Sr_x CuO_4$
the Cu  $1/T_1 T$ increases as T decreases except in a small
region below about T=50K (which we suspect is dominated by superconducting
fluctuations), in contrast to $YBa_2 Cu_3 O_{6.6}$ where the Cu $1/T_1T$ has a
broad maximum at T=150K.  In  $YBa_2 Cu_3 O_{6.6}$ the oxygen $1/T_1 T$ has
nearly the same T-dependence as $\chi_s(T)$ \cite{review,diamag}.

The distinctive behavior of $La_{2-x} Sr_x CuO_4$, namely
a decrease, but {\em not to zero\/}
of $\chi_s$ and a monotonic increase in $1/T_1 T$
as $T~\rightarrow~0$, are
signatures of antiferromagnetism; e.g.
 $\chi_s$ for the 2D S=1/2 Heisenberg model is
maximal at a temperature $T_m$ of order J and
drops by about a factor of two
between $T~\sim~J$ and $T~=~0$ \cite{heis}. However the behavior of
$La_{1.86} Sr_{0.14} CuO_4$ cannot be
interpreted in this way:  although $\chi _s(T)$ is at least qualitatively
consistent
with that of the 2D S=1/2 Heisenberg model with $J$ of $\approx$ 300 - 400K,
the
$1/T$ behavior of the Heisenberg model at $T > J$ is not
observed at $T > 400K$ \cite{batlogg}.
A "two-fluid" model \cite{Johnston} with
only a small density of Heisenberg spins is
not compatible with the relaxation rate data which imply rapid
T-dependent relaxation of all Cu spins.  However, fluctuating regions of local
spin density wave (SDW) ordering can explain the data.  To see this, note
mean field theory predicts a transition at a temperature
$T_{MF}$ to a phase with nonzero staggered magnetization
$\vec N$. In two spatial dimensions,
thermal fluctuations prevent long range order for $T~>~0$ \cite{Hohenberg};
for $T~<~T_{MF}$ the appropriate picture is of
slowly fluctuating domains, with $\vec N$ non-zero, but random from domain to
domain.
We believe $\chi_s$ in this situation may be reasonably approximated by
rotationally averaging the mean field result (which
depends on the angle between the field and $\vec N$),
and rounding out the singularity at $T_{MF}$. This leads to a
susceptiblity which drops by a factor of 2/3 between $T_{MF}$ and 0.
This calculation neglects quantum effects which would
reduce $\chi_s$ further.  These will be less  important
as $T_{MF}$ is decreased, because a lower $T_{MF}$
implies a larger bare correlation length at $T_{MF}$ and
hence a large ``effective spin'' which is ordering.  These arguments
imply that antiferromagnetic fluctuations at $T<T_{MF}$ lead to a $\chi_s(T)$.
However, a mode-coupling analysis of antiferromagnetic fluctuations at
$T>T_{MF}$ yields
a negligible T-dependence of $\chi_s$ \cite{unpub}.
Thus we propose that in the $La_{1.925} Sr_{0.075} CuO_4$ sample the
$T_{MF}$ is rather above room temperature while in the
$La_{1.86} Sr_{0.14} CuO_4$ sample it is somewhat below.  This may be
consistent with
neutron scattering experiments on $La_{1.86} Sr_{0.14} CuO_4$
 \cite{Aeppli} in which quite sharp peaks
are observed emerging
at low T.
 For still
larger Sr concentrations, $T_{MF} < T_c$ and $\chi _s$ displays a very weak
T dependence.

The Cu NMR relaxation is dominated by the relaxational
dynamics of the AF fluctuations, so
$1/T_1 T \sim \sum_q  \xi^{\eta+z} f(q \xi )/ g (q \xi ) \sim \xi^{\eta +z-2}$
where $\xi$ is the correlation length, $\eta$ and $z$ are
critical exponents and f and g are scaling functions for the staggered
susceptiblity and  spin fluctuation energy scale respectively.  Note this form
is not multiplied by $\chi _s(T)$ in contrast to ref \cite{Monthoux}.
The ``MMP'' form proposed \cite{MMP} for $YBa_2 Cu_3 O_7$
corresponds to $\eta~=~z~=~2$,
$f~=~g^{-1} ~=~(1+q^2 \xi^2 )$.
For $T~\ga~T_{MF}$, the  MMP forms are appropriate;
as $T~\rightarrow~0$ the function must cross over to the 2D AF
scaling forms, where $\eta~=~2$ and $z~=~1$.
 The oxygen relaxation rate, however, is due to small q spin
fluctuations, i.e. to fluctuations of the (nearly) uniform magnetization $M_q$
\cite{review}. In an ordered antiferromagnet relaxation is due to spin waves;
At low T the number of thermally excited spin waves is small and the projection
of these onto $M_q$ vanishes as $q~\rightarrow~0$, so that the oxygen
$1/T_1T~\sim T^3$ \cite{Nejat}.  Similarly in a 2d SDW below $T_{MF}$ the
formation of antiferromagnetic domains will lead to an oxygen $1/T_1T$ which
drops
more rapidly than $\chi _s (T)$, in contrast to the prediction in
\cite{Monthoux}.  If our analysis of $\chi _s (T)$ is accepted
then this behavior has already been observed \cite{reven,russ}.

Now consider the $YBa_2 Cu_3 O_{7- \delta}$ system, where
the basic structural unit is a pair of $CuO_2$
planes, separated from the next pair of $CuO_2$
planes by the relatively inert CuO chains \cite{Siegrist}.
We model the spin dynamics of this system using two coupled planes of
antiferromagnetically correlated spins.
This is a gross oversimplification of the physics
of the real material, because it omits the charge
degrees of freedom.We discuss this further below.
The Hamiltonian is
\begin{equation}
H =
J_1 \sum_ {<i,j>,a} \vec S_{i(a)} \cdot \vec S_{j(a)}
+ J_2 \sum_i \vec S_{i(1)} \cdot \vec S_{i(2)}
  \label{Hamiltonian}
\end{equation}

Here $i$ and $j$ label nearest neighbor sites in a
two dimensional square lattice and $a\in\{ 1,2 \}$,  labels
the two different planes.
This model has
a $T~=~0$  phase transition
between a large $J_2$
singlet state and a small $J_2$ antiferromagnetic state.
To study the regime near the transition we use the Schwinger boson
method \cite{Arovas} .  In the model of Eq. \ref{Hamiltonian}
we find the second order
 transition of interest here to be preempted by a first order transition.
  For our detailed calculations we used a slight
variant of Eq. \ref{Hamiltonian}
in which the first order transition is suppressed.
In the mean-field analysis, sums over the momentum q occur;
these may be replaced by an integral over an energy times a density
of states which, for the model of Eq. 1, is constant
near the band edges and logarithmically divergent at band center.  Replacing
this density of states by a constant yields a model with a second order T=0
transition (of the 3D Heisenberg universality class)
at $J_2^* ~=~ 4.48 J_1$.
Insulating $YBa_2 Cu_3 O_{6.0}$ presumably has
$J_1 \gg J_2$
\cite{Tranquada et al.}.  We believe that as the doping increases, the
effective
$J_1$ decreases and $J_2$ increases.  In $YBa_2 Cu_3 O_{6.6}$
antiferromagnetic correlations between pairs of planes have been observed
to persist up to room temperature \cite{Tranquada}.

We have computed the temperature
dependence of  $(\chi _s(T))$ and of the oxygen, yttrium
and copper NMR relaxation rates for various $J_2 > J_2^* $
using  Eq. \ref{Hamiltonian},  the constant density of states,
and a simplified version of the standard NMR form factors
\cite{MMP} in which the Cu transferred hyperfine coupling $B$ was
set to 0.
Some results are shown in Fig.\ 3 for $J_2-J_2^*=0.3$ ; the resemblance
of the curves for Cu and O
to the data for $YBa_2 Cu_3 O_{6.6}$ is evident.
Note in particular existence of two temperature
scales; a higher one, of order $J_2$, at which $\chi _s$ and the oxygen
$1/T_1T$
begin to drop, and a lower scale of order
($J_2 - J^* ) ~<~J_2$ at which the
Cu $1/T_1T$ begins to drop.
  The different T dependences of the Cu and O $1/T_1T$ in our model are due to
both a growing correlation length
and different spin gaps in different regions of q-space, in contrast to the
previous model \cite{MPT}  in which the effects of the spin gap were modelled
by multiplying the MMP
$\chi ^"$ by $\chi _s(T)$, and in which the difference was due only to a T
dependent
correlation length. The difference between the O and Y relaxation rates arises
in our model because the Y nucleus is relaxed only by fluctuations symmetric
under interchange of the two planes; these are the most strongly suppressed by
the tendency to form singlets.  The decrease of the yttrium $1/T_1T$ relative
to $\chi _s$ comes from the phase space restriction involved
in scattering from one small q low-energy state to another and may be a special
feature of this spin-only model. The difference between the O and Y
relaxation rates is not consistent with published Y relaxation rate data
\cite{yT1} but the uncertainty in the Y relaxation rate is large.

 We have also computed the static susceptibitlity at $\vec q=(\pi , \pi )$,
$\chi _{AF}(T)$, finding it to be only very weakly T-dependent for
$J_2-J^*=0.3$, because of the competition between singlet formation (leading to
a $\chi _{AF}$ decreasing
with decreasing T) and antiferromagnetism (leading to the opposite).  For all
$J_2>J_2^*$ we find $\chi _{AF}$  decreases with T for T less than
the temperature at which $1/T_1T$ for Cu has its maximum.  This follows from a
Kramers-Kronig argument: a decrease in $1/T_1T$ implies a
shift of spectral weight in $\chi ''(q,\omega )$ from lower to higher
$\omega$; $\chi '(q,\omega =0) = \int d\omega \chi ''(q,\omega )/\omega$
must then decrease.  It is not consistent with a
recent $T_2$ measurement on $YBa_2Cu_4O_8$ \cite{yast2}.  The $\chi _s$
and Cu $1/T_1T$ for $YBa_2Cu_4O_8$ are similar to those of $YBa_2Cu_3O_{6.6}$
but the
$T_2$ measurement implies that $\int dq [\chi '(q)]^2 \sim \chi_{AF}$ increases
monotonically and smoothly by a factor of 2 between 300K and 100K.
  If confirmed this would imply that as T decreases in $YBa_2Cu_4O_8$  spin
fluctuation weight is not only pushed away from
low frequencies but also pulled down from high frequencies.
Our simple
model does not contain this physics.

A realistic theory must incorporate itinerant carriers, and as in
$La_{2-x}Sr_xCuO_4$ a fermi-surface-instability
description of the magnetic dynamics is required.  One possibility,
a model of two
planes of fermions,
 with direct hopping from plane to plane forbidden, was discussed in ref
\cite{lev}.  Using a gauge theory formalism it was shown that
an arbitrary weak interaction $J_2$ eof Eq.\ 1 leads to a BCS
instability to a ``superconducting'' state in which
every Cooper pair has one member in each plane.
We believe this superconducting state is a mathematical
representation of the between-planes singlet; as pointed out in \cite{lev} it
need not imply the presence of true superconducting order.  We have not yet
incorporated antiferromagnetism in the formalism of
Ref. \cite{lev}, but have shown that direct between-planes hopping of electrons
acts as a pairbreaker of strength proportional to the charge carrier density
\cite{unpub}.
This provides a possible explanation
of the difference between
$YBa_2 Cu_3 O_{6.6}$ and
$YBa_2 Cu_3 O_7$: in the
latter material the much larger hole
density permits interplane hopping which
is strong enough to destroy the interplane pairing.

In this paper we have proposed explanations for the
spin gap behavior observed in underdoped cuprates.
A crucial datum for our interpretation is the $T~\rightarrow~0$
extrapolation of $\chi_s (T)$.
We have argued that it is non-zero and indeed large in
$La_{2-x} Sr_x CuO_4$ .
If it is small, then a one-plane quantum disordered
phase must be considered for
$La_{2-x} Sr_x CuO_4$, and the evidence
for interplane pairing is in
$YBa_2 Cu_3 O_{6.6}$ is weakened.  Two important consequences are: (a) in
$La_{2-x}Sr_xCuO_4$ samples with a T-dependent $\chi_s$ the oxygen $1/T_1T$
should decrease more rapidly than $\chi_s$ with decreasing T and (b) in
$YBa_2Cu_3O_{6.6}$ the Y relaxation rate should drop more rapidly than the O
relaxation rate as T decreases.
\newpage
\section*{Acknowledgments}
A. J. M. thanks B. Batlogg and R. E. Walstedt for helpful
discussions and the generous sharing of unpublished data and C. Berthier
for drawing his attention to Ref. \cite{yast2} and helpful discussions.  One of
us (H. M. ) thanks A.T.\&T. Bell Laboratories
for hospitality while part of this work was performed.  H. M. was supported
by NSF grant NSF PHY89-04035.

\newpage

\section*{Figure Captions}

\noindent
{\bf Fig. 1.:}  Spin suspectiblities of high $T_c$ materials obtained  from
data as described in the text, along with the theoretical susceptiblity of the
2d S=1/2 heisenberg
model \cite{qheis}. The powder average of the observed $\chi_b$ for $La_2CuO_4$
at
at $T~=~750$K  is within 0.2 states/(ev-Cu) of the theoretical result.

\vskip0.3cm
\noindent
{\bf Fig. 2.:} Copper NQR relaxation rates of high-$T_c$ materials,from Refs.
\cite{LaSrT1,YBaT1}.   Ref. \cite{YBaT1} used a
normalization convention which differs from one used here by a factor
of 3.  We have reexpressed the data of ref \cite{YBaT1} accordingly.

\vskip0.3cm
\noindent
{\bf Fig. 3.:} Copper, oxygen  and yttrium relaxation rates
 calculated for model of two coupled antiferromagnetically correlated planes
using Schwinger boson mean field analysis of eq 1 for
$J_2-J^*=0.3$.  The left ordinate shows the Cu and O relaxation rates $1/T_1T$
(solid lines); the right ordinate shows the ratio of the O and Y $1/T_1T$ to
the calculated spin suceptibility $\chi_s$.

\end{document}